\documentclass[reprint,eqsecnum,floats,aps,amsmath,amssymb,nofootinbib,prd,onecolumn, showpacs]{revtex4-2}
\usepackage{graphicx,physics}
\usepackage{amsmath,amssymb,mathtools,mathrsfs}
\usepackage{hyperref}
\usepackage{graphicx}
\usepackage{subfigure}
\usepackage{arydshln}
\usepackage{xcolor}
\usepackage{braket}
\usepackage{tensor}
\usepackage{enumitem,array,textcomp}
\usepackage[utf8x]{inputenc}

\DeclareMathOperator\arctanh{arctanh}

\begin{document}
	
	\title{Two-time alternative to the Ashtekar-Olmedo-Singh black hole interior}
	\author{Alejandro Garc\'ia-Quismondo}
 	\email{alejandro.garcia@iem.cfmac.csic.es}
	\affiliation{Instituto de Estructura de la Materia, IEM-CSIC, Serrano 121, 28006 Madrid, Spain}
	\author{Guillermo  A. Mena Marug\'an}
	\email{mena@iem.cfmac.csic.es}
	\affiliation{Instituto de Estructura de la Materia, IEM-CSIC, Serrano 121, 28006 Madrid, Spain}
	
\begin{abstract}
We investigate the viability of a recently proposed generalization of the Ashtekar-Olmedo-Singh spacetime for the effective description of the interior region of a Schwarzschild black hole within the framework of loop quantum cosmology. The approach is based on a choice of polymerization parameters that is more general than the ones previously considered in the literature and that results in the natural appearance of two times to describe the solutions. If one is interested in examining the physics derived from this model, it is fundamental to ensure that one can attain a well-defined effective geometry in the whole region under consideration, in particular as regards the redundancy of the two times, which one needs to express in terms of a single time coordinate. In order to determine whether this requirement is met, we analyze the definition of these times and their relation. We show that one can reach an acceptable interior spacetime geometry by exploiting the freedom to define the origins of the two times independently.
\end{abstract}

\pacs{98.80.Qc, 04.70.Dy, 04.60.Ds, 04.60.−m.}
	
\maketitle					

\section{Introduction}\label{sec:intro}

General relativity is one of the cornerstones of modern theoretical physics and has reshaped our comprehension of the world \cite{Wald,Hawking-Ellis}. It provides a framework to describe the physics of gravitational systems in planetary and cosmological scales, agreeing with experiments and observations to high levels of precision. Nevertheless, it is believed to be incomplete for a number of reasons. On the one hand, it leads to the prediction of singularities \cite{Hawking-Ellis}, where the theory ceases to be applicable. These singularities arise in a wide variety of scenarios, many of which are of physical significance. On the other hand, it is not compatible with the principles of quantum mechanics \cite{CT,Galindo}, which lie at the very heart of the modern descriptions of elementary interactions and matter. This is in direct tension with the basic observation that matter gravitates, which may suggest that the quantum cannot be ignored if one wishes to achieve a complete description of gravitational physics. The widespread belief is that the incorporation of the ideas of quantum mechanics into our understanding of gravity would not only resolve this apparent tension, but also cure the singularities that plague general relativity. For this reason, there has been a collective effort to bring together the principles of general relativity and quantum mechanics, effort which already existed in embryo in the early days of general relativity and is still ongoing today. Naturally, a number of different approaches have been adopted, leading to diverse proposals to formulate a candidate theory of quantum gravity (see, e.g., Refs. \cite{ALQG,Thiem,Mod,ST,CDT}). 

Loop quantum gravity (LQG) stands as one of the most promising of such proposals \cite{ALQG,Thiem}. It is a background independent, non-perturbative quantization of general relativity in 3+1 dimensions. In its canonical form, it is based on a choice of fundamental variables, given by the holonomies of the Ashtekar-Barbero connection along loops and the fluxes of densitized triads through surfaces, and on the selection of a quantum representation of the holonomy-flux algebra compatible with background independence, which turns out not to be unitarily equivalent to the Fock representation of standard quantum field theory. Although the quantization program of LQG remains unfinished, the study of highly symmetric systems, which provide a natural arena to explore the effects of quantum gravity, has undergone a rapid development within this quantum framework. The research field born from the application of LQG techniques to systems displaying a large number of symmetries (such as cosmological and black hole spacetimes) is commonly referred to as loop quantum cosmology (LQC) \cite{AS,LQCG}. In the cosmological front, the application of techniques inspired in LQG has been successful at obtaining a complete description of a variety of cosmologies, ranging from homogeneous and isotropic ones \cite{APS2,MMO} to models believed to be fit to describe the early stages of our Universe \cite{dressed,CMBdressed,hybridGIper,CMBhybrid}. As far as black holes are concerned, there is also a rich literature \cite{1,2,3,18b,4,5,6,7,8,9,10,11,12,13,14,15,16,17,18,18b,19,20,21,22,23} about the application of LQG\footnote{The references in this paragraph do not intend to provide an exhaustive bibliographic list. To get a more accurate picture of the extent of the field of LQC, an appropriate starting point may be Refs. \cite{ALQG,AOS,AOS2} and the references cited there.}. 

Recently, a new proposal for the effective\footnote{In this paper, we will use the term \emph{effective} to characterize a description, model, or mathematical object that is inherently classical but incorporates corrections of quantum geometrical origin.} description of Schwarzschild black holes in LQC has been put forward by Ashtekar, Olmedo, and Singh (AOS) \cite{AOS,AOS2,AO}. This effective model is especially designed to cope with the loop quantum corrections to the geometry of black holes with masses much larger than the Planck mass. It has attracted a fair degree of attention owing to the claims that it is free from some of the pathological properties present in previous related works (e.g., the dependence on fiducial structures or the appearance of some quantum geometry effects in regions of low spacetime curvature). This model is based on a very particular choice of the two polymerization parameters that regulate the introduction of quantum effects in the system: they are selected in such a way that they are \emph{constants of motion} but \emph{not} constant on the whole phase space. This approach turns out to replace the classical central singularity with a transition surface serving as the future boundary of a trapped region and as the past boundary of an anti-trapped one. Furthermore, an additional boundary is found beyond the transition surface, which is interpreted to be a white hole horizon. Therefore, the physical picture derived from this model appears to be such that the classical Schwarzschild interior is effectively extended to encompass a region bounded to the past by a black (hole) horizon and to the future by a white (hole) horizon,\footnote{Our convention to distinguish between future and past is closely related to the standard notion of light cones pointing radially inward in the interior of a Schwarzschild black hole, with the central singularity becoming the future endpoint of all timelike geodesics. This concept suggests that the future corresponds to decreasing values of the coordinate time introduced in Sec. \ref{sec:basics}.} where the effective spacetime metric is smooth and the curvature invariants derived from it are finite (and, in fact, bounded above by quantities that are independent of the mass of the black hole under consideration). Moreover, the effective model has been brought to completion by extending it to describe the exterior region as well, leading to a geometry that can be smoothly joined with the interior solution both to its past and its future, resulting in an effective extension of the totality of the Kruskal spacetime. 

In spite of its attractive features from the point of view of the singularity resolution, the model seems to be not without problems. It has been pointed out that the effective exterior metric proposed in Refs. \cite{AOS,AOS2} does not display a standard behavior at spatial infinity. Indeed, the exterior geometry turns out to be asymptotically flat just in an elementary sense, since it has been argued that it can be conformally related to a metric that contains a deficit solid angle \cite{Bouhm} (see Ref. \cite{AO} for a complementary viewpoint on this issue). The appearance of this deficit might be connected with claims that an effective metric such as the AOS one cannot be derived from a loop quantization that preserves strict covariance while respecting spherical symmetry \cite{bojosb}.  
Additionally, according to Ref. \cite{N}, the way in which the polymerization parameters were originally treated seems to ignore the hypothesis that they are constants of motion. In this respect, in Refs. \cite{AOS,AOS2}, Ashtekar, Olmedo, and Singh employed an argument based on an extension of the phase space in order to support their approach, in which the parameters are handled as constants in the Hamiltonian derivation of the equations of motion. Nevertheless, the authors of Ref. \cite{N} have indicated that this makes the relation between the proposed Hamiltonian and the dynamical equations unclear, given that an extra phase space dependent factor would enter the equations of motion should the non-trivial nature of the polymerization parameters be taken into account, leading to a more involved dynamics \cite{N,AG}. 

In view of this situation, and focusing exclusively on the Hamiltonian derivation of the AOS solution, an alternative approach has been proposed in Ref. \cite{AG} to obtain the dynamical equations while considering the non-commutativity of the polymerization parameters with the canonical variables, bringing together a treatment of these parameters as true constants of motion and the undoubtedly interesting physical results of the original AOS model. In that paper, we introduced an alternative prescription for the selection of the polymerization parameters that extends the ideas of Ref. \cite{N}. Supported on considerations of generality, we have suggested that one should allow that the parameters capture phase space contributions coming from two separate sectors that had been viewed as decoupled in the literature previous to our proposal. This leads to a richer variety of dynamical equations, which differ from those obtained by considering constant parameters in a pair of multiplicative phase space dependent factors, as it is also the case of Ref. \cite{N} (although the factors we found are considerably more complicated owing to the coupling between sectors). After reabsorbing those factors via time redefinitions and obtaining the form of the dynamical solutions, we discussed in Ref. \cite{AG} some consequences of the relation between the newly-defined times and their behavior in the asymptotic limit of infinitely large black hole masses. The effective spacetime geometry derived with this procedure is fundamentally different from that in the original works. This fact has left the door open to a possible alleviation of the problems of the model. The objetive of the present work is to develop the preliminary ideas introduced in Ref. \cite{AG} and, in particular, to address the issue of whether there exists any obstruction to our proposal such as it was originally formulated, with special emphasis on a good behavior of all the interior geometry. This step is vital if we want to examine the physical properties of the model at a later point, in order to fully comprehend the effective theory and set it on a firm foundation on top of which one may complete the quantization program. Therefore, in this article we concentrate all our attention exclusively on this issue, which we consider prior to any investigation of an algebraic or canonical quantization of the model and of the subsequent quantum properties of the system. 

The article is structured as follows. First, we introduce the basics of the effective model under consideration in Sec. \ref{sec:basics}. Then, in Sec. \ref{sec:obstructions} we proceed to the investigation of possible obstructions to the presented formalism. More precisely, this contains: (1) the study of the integrability of $F_{ij}$ and the invertibility of $G_i$ (for a precise definition of these functions, see Sec. \ref{sec:obstructions} or Ref. \cite{AG}), performed in Sec. \ref{subsec:integrabilityinvertibility}; (2) a discussion about the image of $G_i$, which can be found in Sec. \ref{subsec:imageG}; and (3) an analysis of the properties of the factors $C_{ij}$ along dynamical trajectories, contained in Sec. \ref{subsec:analysisC}. In Sec. \ref{sec:redefinitionorigins}, we consider the possibility of independently redefining the origin of one of the times of the system and evaluate the impact of such a redefinition on the viability of the model. Finally, we summarize the main results and discuss their consequences in Sec. \ref{sec:conclusion}.

Throughout this article, we will use the following notation. Letters from the middle of the Latin alphabet ($i$, $j$, $k$...) are used as phase space labels, taking the value $b$ or $c$ depending on whether they refer to the radial or the angular phase space sector, respectively. Unless explicitly stated otherwise, if two such labels appear in any given expression, they are assumed to be different from each other. Letters from the middle of the Greek alphabet ($\mu$, $\nu$...) denote spacetime indices, with values ranging from 0 to 3. Finally, we adopt natural units, setting the speed of light and the reduced Planck constant to one.  

\section{The model} \label{sec:basics}

We start by briefly introducing the main ideas of our proposal following Ref. \cite{AG}, where they were first put forward. From now on, we will focus our attention on the study of the interior region of a non-rotating, uncharged black hole. This region admits a foliation in homogeneous, spacelike Cauchy hypersurfaces, namely those characterized by a constant value of the Schwarzschild coordinate $r$. This property, which is not exhibited by the exterior region, allows for the construction of a Hamiltonian description of the system in terms of a finite dimensional phase space. Indeed, after the imposition of the Gauss constraint arising as a result of employing a triadic formulation, all dynamical information about the Ashtekar-Barbero variables is encoded in two canonical pairs, $(b,p_b)$ and $(c,p_c)$ \cite{AOS,AOS2}. The first pair (comprised of the connection variable $b$ and the triad variable $p_b$) refers to the radial sector of the spatial hypersurfaces, whereas the other pair (also composed by a connection variable $c$ and a triad variable $p_c$) is related to the angular degrees of freedom. In view of this distinction (and using a terminology motivated by our previous comments), it is useful to differentiate between what we will call the \emph{radial sector} and the \emph{angular sector} of phase space, to which we will often refer throughout this article. The non-vanishing Poisson brackets of these fundamental variables are
\begin{align}
\{b,p_b\}=G\gamma,\qquad
\{c,p_c\}=2G\gamma,
\end{align}
where $G$ is the Newton constant and $\gamma$ is the Immirzi parameter. In terms of these canonical variables, the spacetime line element can be written as
\begin{align}
ds^2=g_{\mu\nu}dx^\mu dx^\nu=-N^2dt^2+\dfrac{p_b^2}{L_o^2|p_c|}dx^2+|p_c|d\Omega^2,\label{effmetric}
\end{align}
where $g_{\mu\nu}$ is the spacetime metric, $N$ is the lapse function, $x$ is a radial coordinate in the interior region, $L_o$ is a fiducial length associated with this coordinate (hence, physical results must have a well-defined limit when $L_o\to\infty$), and $d\Omega^2=d\theta^2+\sin^2\theta d\phi^2$ is the metric of the unit 2-sphere in terms of the polar and azimuthal angles, $\theta$ and $\phi$. 

As a consequence of the symmetries of general relativity, the above canonical variables must satisfy certain constraints. In fact, given that the freedom associated with the Gauss constraint has already been fixed in the process of defining the canonical variables and the spatial diffeomorphism constraint is trivial in the chosen foliation, the only remaining non-trivial constraint is the effective Hamiltonian $H_{\rm eff}$ itself, which generates time reparametrizations and must vanish on the dynamical solutions. For the choice of lapse function
\begin{align}
N=\dfrac{\gamma \delta_b\sqrt{|p_c|}}{\sin\delta_bb},
\end{align} 
associated with a time $t$ that we will call \emph{coordinate time}, the product with the effective Hamiltonian constraint turns out to be \cite{AOS,AOS2}
\begin{align}
N H_{\rm eff}&=\dfrac{L_o}{G}(O_b-O_c),\\
O_b&=-\dfrac{1}{2\gamma}\left(\dfrac{\sin\delta_bb}{\delta_b}+\dfrac{\gamma^2\delta_b}{\sin\delta_bb}\right)\dfrac{p_b}{L_o},\label{Ob}\\
O_c&=\dfrac{1}{\gamma}\dfrac{\sin \delta_cc}{\delta_c}\dfrac{p_c}{L_o},\label{Oc}
\end{align}
where $\delta_b$ and $\delta_c$ are the two polymerization parameters that regulate the introduction of quantum effects in the system. Indeed, in the limit where both parameters vanish, the above Hamiltonian reduces to that of general relativity written in terms of our variables. Notice that, with this choice of lapse, the Hamiltonian constraint presents a remarkably simple structure: up to a multiplicative constant, it is given by a difference of two quantities, $O_b$ and $O_c$, which only depend on the canonical variables of either the radial or the angular sector of phase space and their respective polymerization parameter. Therefore, it is immediate to realize that, unless the polymerization parameters introduce a cross-dependence between the radial and angular sectors, the dynamics of the two sectors is decoupled and, up to constants, $O_b$ and $O_c$ generate their respective dynamics. For this reason (and although this interpretation only holds when no cross-dependence is introduced), we will refer to $O_b$ and $O_c$ as the radial and the angular \emph{partial Hamiltonians}, respectively.

The next step is the selection of the polymerization parameters, $\delta_b$ and $\delta_c$. Although several approaches have been explored in the literature, we will focus our attention on definitions such that they are constants of motion, i.e., constant along dynamical trajectories but not on the whole of the phase space. However, this restriction leaves an ample freedom and we need to adopt a concrete prescription. The proposal of the authors of the original model is based on the observation that the partial Hamiltonians are in fact constants of motion themselves. Not only that, given the form of the Hamiltonian constraint, it follows that both partial Hamiltonians have the same on-shell value, which we call $m$. This quantity has been found to be related to the mass of the black hole \cite{AOS,AOS2}. In light of these considerations, the AOS proposal suggests that the parameters be taken as functions of the constant of motion $m$, and thus of the black hole mass, supporting this choice with an argument that appeals to an extension of the phase space (for more details, we encourage the reader to consult Refs. \cite{AOS,AOS2}). In a later work \cite{N}, it has been argued that such an approach is not consistent with the premise of selecting constants of motion as polymerization parameters. Instead, it has been proposed that each parameter be treated as a function of its associated partial Hamiltonian and, then, account for the non-vanishing Poisson brackets of these parameters in the derivation of the equations of motion, a procedure that leads to dynamical equations that differ from those considered in the original works. In this way, one would incorporate the non-constant nature of the polymerization parameters while still ensuring that they are functions of $m$ on the constraint surface.

In an attempt to combine the positive aspects of these two approaches, we put forward an alternative proposal that incorporated the ideas of Ref. \cite{N} but tried to reconcile them with the interesting properties of the original model, and in particular the displayed singularity resolution \cite{AOS,AOS2,AO}. We argued that, since both partial Hamiltonians have the same on-shell value, one ought not to be able to distinguish their individual on-shell contributions. Therefore, the most general choice should be one of the type
\begin{align}
\delta_i=f_i(O_b,O_c),\label{deltaiObOc}
\end{align}
where we have used a compact notation to refer to both polymerization parameters by means of a subindex $i=b,c$ (see the last paragraph of Sec. \ref{sec:intro} for a more detailed comment on this notation). It is important to emphasize that these definitions introduce a cross-dependence in the Hamiltonian that in principle breaks the decoupling of the radial and angular sectors of phase space. In the absence of this decoupling, which is present in all other related works on the model, the resulting dynamical equations are \cite{AG}
\begin{align}
\partial_t i&=C_{ij}\left[s_i\dfrac{L_o}{G}\{i,p_i\}\dfrac{\partial O_i}{\partial p_i}\right],\label{partialti}\\
\partial_t p_i&=C_{ij}\left[-s_i\dfrac{L_o}{G}\{i,p_i\}\dfrac{\partial O_i}{\partial i}\right],\label{partialtpi}
\end{align} 
where $i$ and $j$ are assumed to be different and $s_i$ is a sign defined as follows:
\begin{align}
s_i=\left\{\begin{matrix}
+1&\textrm{if } i=b,\\-1&\textrm{if } i=c.
\end{matrix}\right.
\end{align}
It is straightforward to realize that the objects in square brackets in Eqs. \eqref{partialti} and \eqref{partialtpi} are nothing but the equations of motion that would be obtained if the polymerization parameters were treated as constants. Therefore, all the information about the non-trivial nature of the parameters as functions on phase space is enclosed in the factors $C_{ij}$, which are\footnote{Should the functions $f_b(O_b,O_c)$ and $f_c(O_b,O_c)$ be fixed (ideally from first principles or, alternatively, from empirical inputs), these phase space dependent factors would be totally determined. Nonetheless, in this paper we adopt a more modest approach and simply impose a minimum set of conditions that these factors must satisfy to allow us develop our analysis, namely, in principle that they are finite and such that, in the limit of infinite mass, the polymerization parameters coincide on shell with those in Ref. \cite{AOS,AOS2}. See the rest of the main discussion for further details.}
\begin{align}
C_{ij}=\dfrac{1-\Delta_{jj}-\Delta_{ji}}{(1-\Delta_{ii})(1-\Delta_{jj})-\Delta_{ij}\Delta_{ji}},\qquad \Delta_{ij}=\dfrac{\partial O_i}{\partial \delta_i}\dfrac{\partial f_i}{\partial O_j},\label{Cij}
\end{align}
where the subindices of $\Delta_{ij}$ are allowed to be equal and $(1-\Delta_{ii})(1-\Delta_{jj})-\Delta_{ij}\Delta_{ji}$ is assumed to be different from zero in the derivation of the equations of motion. According to our conventions, our choice of parameters is then reflected in the presence of two phase space dependent factors, $C_{bc}$ and $C_{cb}$, that appear multiplicatively in the dynamical equations. The fact that they show up precisely in this way implies that they can be reabsorbed through appropriate time redefinitions, one per sector of phase space. Indeed, if we consider a \emph{radial time} $t_b$ and an \emph{angular time} $t_c$ given by
\begin{align}
dt_i=C_{ij}dt,\label{defti}
\end{align}
the dynamical equations reduce to those resulting from parameters that are constants on the whole phase space. Thus, we attain a set of equations of motion that is identical in form to that of Refs. \cite{AOS,AOS2} except for the fact that they are written in terms of two newly-defined times instead of a single one\footnote{Although these time redefinitions prove useful in the case under consideration, we will require that the system be expressible in terms of a single time as a necessary condition to obtain a well defined effective geometry. Note that this does not imply that a similar situation with multiple times necessarily arises in other less symmetrical scenarios in LQC. For instance, in the case of Bianchi I cosmologies, the regularization of the Hamiltonian involves the introduction of three phase space dependent length parameters (one per direction of the spatial sections), but there is no need of introducing three different times in that setting \cite{BI}.}. We can proceed to integrate these equations, getting solutions that are formally identical to those obtained in Refs. \cite{AOS,AOS2} but expressed in terms of the two new times. We find
\begin{align}
\tan\dfrac{\delta_c c(t_c)}{2}&=\dfrac{\gamma L_o \delta_c}{8m}e^{-2t_c},\\
p_c(t_c)&=4m^2\left(e^{2t_c}+\dfrac{\gamma^2L_o^2\delta_c^2}{64m^2}e^{-2t_c}\right),\\
\cos\delta_bb(t_b)&=\dfrac{1+b_o\tanh \dfrac{b_o t_b}{2}}{1+b_o^{-1}\tanh \dfrac{b_o t_b}{2}},\\
p_b(t_b)&=-2L_o m\dfrac{\sin\delta_bb(t_b)}{\gamma \delta_b}\dfrac{1}{1+\dfrac{\sin^2\delta_bb(t_b)}{\gamma^2\delta_b^2}}\label{pbeff},
\end{align}
where $b_o=\sqrt{1+\gamma^2\delta_b^2}$ and, following the conventions of Refs. \cite{AOS,AOS2,AO}, $b>0$, $p_b\leq 0$, $c>0$, and $p_c\geq 0$. Additionally, the way in which the integration constants are fixed in those references implies that the horizon lies at $t_b=0$, instant at which both the connection and triad variables of the radial sector vanish. In principle, we choose the origin of $t_c$ (i.e. $t_c=0$) on the horizon as well. This facilitates the comparison of our arguments and results with previous works. In Sec. \ref{sec:redefinitionorigins}, we will return to this issue and comment on the available freedom of choice, which will play an important role in our discussion. It is worth pointing out that the effective solutions display reflection symmetries, one per sector of phase space. Indeed, it is straightforward to see that the angular partial Hamiltonian $O_c$ is left invariant by the transformation $t_c\to \ln(\gamma L_o\delta_c/8m)-t_c$, which leaves $p_c$ unchanged and takes $\delta_cc$ to $\pi-\delta_cc$. There is a completely analogous situation in the radial sector under the transformation $t_b\to -(4/b_o)\arctanh (1/b_o)-t_b$. 

In the classical limit where $\delta_i\to 0$, the above solutions reduce to the ones found in general relativity. Since the radial and angular times coincide in such a classical limit \cite{AG} and are equal to the coordinate time [as can be inferred from their definitions \eqref{defti}, together with the fact that $C_{ij}\to1$ in the considered classical limit], a straightforward computation \cite{AOS,AOS2} shows that the corresponding classical solutions are given by\footnote{We note that the limit of vanishing mass of Eq. \eqref{classicalc} is not well defined. This is not surprising since, in that limit, the interior of the black hole disappears and there exists no classical Kantowski-Sachs solution describing it.} 
\begin{align}
c(t)&=\dfrac{\gamma L_o}{4m}e^{-2t},\label{classicalc}\\
p_c(t)&=4m^2e^{2t},\\
b(t)&=\gamma \sqrt{\abs{e^{-t}-1}},\\
p_b(t)&=-2L_o m \dfrac{\sqrt{\abs{e^{-t}-1}}}{1+\abs{e^{-t}-1}}.
\end{align}
It is immediate to see that the phase space counterpart of the classical central singularity is found in the limit $t\to-\infty$, where the triad variables vanish and the connection variables diverge. Therefore, the interior region of the Schwarzschild black hole corresponds to the coordinate time interval $(-\infty, 0)$.

In the effective theory where the polymerization parameters are different from zero, we can try and identify the time intervals corresponding to the interior region by similar means. However, as a result of the inclusion of quantum effects, there is no singularity in the sense that neither the momentum variables vanish nor the connection variables diverge at any interior point along any dynamical trajectory. In fact, note that the absolute value of the triad variables is bounded from below. In the case of $p_c$, it is clear that it reaches a local minimum when $t_c$ equals a critical value
\begin{align}
t_c^{\mathcal{T}}=\dfrac{1}{2}\ln \dfrac{\gamma L_o\delta_c}{8m},
\end{align}
which is negative for sufficiently massive black holes (or, in other words, for values of the polymerization parameters that are sufficiently small), which are precisely the ones aimed to be described by the model under consideration \cite{AOS,AOS2}. This critical value defines a spacelike hypersurface $\mathcal{T}$ that we will refer to as the \emph{transition surface}, given its physical interpretation in Refs. \cite{AOS,AOS2,AO}. From the expression of $t_c^\mathcal{T}$, it is straightforward to see that it tends to negative infinity in the classical limit. In this sense, it is often said that the transition surface ``replaces'' the central singularity in the effective theory. Hence, we can conclude that the classical interior region still corresponds to negative values of $t_c$, now within the interval $(t_c^\mathcal{T},0)$ of angular times\footnote{From now on, we will interchange the use of the radial and angular times, depending on which one is more convenient at each point of the discussion. It should be born in mind that Eq. \eqref{defti} can be employed to obtain a dictionary between both times by solving for $dt$ in both expressions, equating the results, and performing an integration. This provides us with an equality of two functions of $t_b$ and $t_c$, respectively, which ought to define an implicit relation between these two times. For further details, see the discussion of Sec. \ref{sec:obstructions}.}. This means that a brand new region of a purely quantum origin appears beyond the transition surface, i.e. for values of the angular time $t_c<t_c^\mathcal{T}$. In order to analyze in a meaningful way how the physical picture is modified as a result of the inclusion of quantum gravitational effects through polymerization parameters that have contributions from both sectors of phase space \eqref{deltaiObOc}, we need to determine first whether the effective spacetime metric can be defined satisfactorily in the totality of the interior region. The study of this issue, which was not addressed in Ref. \cite{AG}, is precisely the aim of Sec. \ref{sec:obstructions}.

\section{Possible obstructions to the model} \label{sec:obstructions}

In order to examine the physical properties of the solution obtained with our proposal, it is necessary to discuss whether there exist obstructions to our two-time formalism in the first place. In other words, we must analyze whether the effective spacetime metric is in fact well defined at every point of the interior region, in the sense that it can always be written in terms of a single time coordinate (by patches, if needed). Conditions that are necessary to have such a well-defined effective metric in a neighborhood of any point in terms of one of our two times are the following:
\begin{itemize}
\item[a)] One of the times $t_b$ or $t_c$ can be reexpressed in terms of the other.
\item[b)] The time component of the effective line element can be rewritten in terms of either $dt_b^2$ or $dt_c^2$, depending on which time can be expressed as a function of the other.
\end{itemize}
Naturally, we need these conditions to hold at every possible point in order to cover the whole interior region. 

From the definitions of the radial and angular times [see Eq. \eqref{defti}], it follows immediately that
\begin{align}
dt^2=\dfrac{dt_b^2}{C_{bc}^2}=\dfrac{dt_c^2}{C_{cb}^2}.
\end{align}
Therefore, the contribution in the line element that is proportional to $dt^2$ can always be rewritten in terms of the square of the differential of the appropriate time variable provided that the inclusion of the corresponding factor $1/C_{ij}^2$ does not lead to singularities. This issue will be studied carefully in Sec. \ref{subsec:analysisC}.

Let us then focus on the other point, namely, whether it is possible to rewrite at least one of the times in terms of the other. This requires an analysis of the implicit relation between the radial and angular times. This relation can be obtained from the equality
\begin{align}
C_{cb}dt_b=C_{bc}dt_c,\label{esta}
\end{align}
which is easily derived from the definitions of $t_b$ and $t_c$. Since the denominator of $C_{ij}$ is symmetric under the exchange of its indices [see Eq. \eqref{Cij}], only the numerators of these factors, 
\begin{align}
 F_{ij}=1-\dfrac{\partial O_j}{\partial \delta_j}\left(\dfrac{\partial f_j}{\partial O_j}+\dfrac{\partial f_j}{\partial O_i}\right),
\end{align}
are relevant for the implicit relation between the radial and angular times (assuming that the denominator of $C_{ij}$ is finite and non-zero, see Sec. \ref{subsec:analysisC}). In this way, after an integration, Eq. \eqref{esta} becomes
\begin{align}
G_b(t_b)=\int_{t_b}^0 F_{cb}(t_b')dt_b'=\int_{t_c}^0F_{bc}(t_c')dt_c'=G_c(t_c),
\end{align}
where the functions $F_{ij}$ must be evaluated on solutions and, hence, on the Hamiltonian constraint surface. If we rewrite the parameters $\delta_i=f_i(O_b,O_c)$ as functions of the linear combinations of the partial Hamiltonians given by $\mu_1=(O_b+O_c)/2$ and $\mu_2=(O_b-O_c)/2$, and provided that the polymerization parameters are at least $\mathcal{C}^1$, we get
\begin{align}
F_{ij}|_{\rm on-shell}=1-\dfrac{\partial f_j(m,0)}{\partial m}\dfrac{\partial O_j}{\partial \delta_j}\big|_{\rm on-shell},\label{Fijonshell}
\end{align}
where we have used that $\mu_1|_{\rm on-shell}=m$ and $\mu_2|_{\rm on-shell}=0$. Additionally, $\partial/\partial m$ denotes the derivative with respect to the quantity $\mu_1$ evaluated on the constraint surface. As a result, the implicit relation between the two times can be recast as
\begin{align}
-G_b(t_b)=t_b-\dfrac{\partial f_b}{\partial m}\int_0^{t_b}\dfrac{\partial O_b}{\partial \delta_b}(t_b')dt_b'=t_c-\dfrac{\partial f_c}{\partial m}\int_0^{t_c}\dfrac{\partial O_c}{\partial \delta_c}(t_c')dt_c'=-G_c(t_c),\label{-Gb=-Gc}
\end{align} 
where we have omitted the on-shell evaluation for the sake of simplicity. As we anticipated, this expression gives an implicit relation between both times on shell, $G_b(t_b)=G_c(t_c)$. We will carry out a detailed analysis of this relation in the following.

It is clear that the need to express the effective metric in terms of a single time in the whole interior region imposes a series of requirements on the functions $F_{ij}$ and $G_i$. In particular, these functions must satisfy the following conditions:
\begin{itemize}
\item[i)] The primitives $G_i$ must exist. This is equivalent to requiring that $F_{ji}$ be integrable.
\item[ii)] At least one of the primitives $G_i$ must be invertible at each point of the spacetime region under consideration. 
\item[iii)] The images of the two primitive functions $G_i$ must coincide in the whole of the interior region.
\end{itemize}
If there is a subregion in which the functions $F_{ij}$ are not integrable or none of the primitives is invertible, there will exist obstructions to define an effective metric in terms of a single time in that part of the interior region. Moreover, if the images of the primitives $G_i$ differ, there exists a subregion where Eq. \eqref{-Gb=-Gc} cannot hold, preventing that the angular and radial times can be related there. 

Let us begin by examining in Sec. \ref{subsec:integrabilityinvertibility} whether the integrability and local invertibility conditions can be satisfied. This analysis will also provide us with valuable tools to address the study of the images of $G_b$ and $G_c$, which we will carry out in Sec. \ref{subsec:imageG}.

\subsection{Necessary conditions for a well-defined and invertible time relation} \label{subsec:integrabilityinvertibility}

Let us consider first the integrability of $F_{ij}$. A direct inspection of Eq. \eqref{-Gb=-Gc} confirms that this condition is equivalent to the integrability of $\partial O_j/\partial \delta_j$. Since any continuous function on a closed interval is integrable on that interval, and $\partial O_j/\partial \delta_j$ is continuous on its domain because it is an elementary function\footnote{In other words, it can be obtained through a finite number of compositions and combinations of the four fundamental operations on basic elementary functions (powers, exponentials, logarithms, and direct and inverse trigonometric and hyperbolic functions).}, both for $j=b$ and for $j=c$, it suffices to verify that this domain always contains the integration interval.

Taking the partial derivatives of $O_b$ and $O_c$ with respect to their associated polymerization parameter, we obtain 
\begin{align}
\dfrac{\partial O_b}{\partial \delta_b}&=-\dfrac{1}{2\gamma }\left(1-\dfrac{\gamma^2\delta_b^2}{\sin^2\delta_bb}\right)\dfrac{\delta_bb\cos\delta_bb-\sin\delta_bb}{\delta_b^2}\dfrac{p_b}{L_o},\label{dObddeltab}\\
\dfrac{\partial O_c}{\partial \delta_c}&=\dfrac{1}{\gamma}\dfrac{\delta_cc\cos\delta_cc-\sin\delta_cc}{\delta_c^2}\dfrac{p_c}{L_o}.\label{dOcddeltac}
\end{align}
For finite values of the polymerization parameters, the domain of $\partial O_c/\partial \delta_c$ as a function of the angular time is the real line $\mathbb{R}$ and, in particular, contains the time interval corresponding to the interior region. The same statement holds as well for $\partial O_b/\partial \delta_b$ as a function of $t_b$, as can be easily verified. Thus, both objects are integrable, a fact which ensures that the primitives $G_i(t_i)$ exist. Obviously, this result does not guarantee that the functions $G_i(t_i)$ can be written in terms of elementary functions, as we will see later on. 

We turn to discuss the invertibility of the primitives $G_i$. The inverse function theorem states that, if a function is differentiable at a given point and its derivative is continuous and non-vanishing at that point, then the function is invertible in a neighborhood of it. Since the functions $-F_{ji}(t_i)$, which are the derivatives of $G_i(t_i)$, are continuous, the theorem ensures that the primitives are locally invertible except around the zeroes of the functions $F_{ji}(t_i)$. 

We note that the discussion of these zeroes will also provide us with valuable information about the behavior and form of the primitives $G_i$. Let us start by studying the zeroes of $F_{bc}(t_c)$. According to Eqs. \eqref{Fijonshell} and \eqref{dOcddeltac}, and taking into consideration that $p_c\sin\delta_cc$ is proportional to $O_c$ and, hence, to $m$ on shell, we have
\begin{align}
F_{bc}(t_c)=1-\dfrac{1}{\gamma L_o\delta_c^2}\dfrac{\partial f_c}{\partial m}\left[\delta_cc(t_c)\cos\delta_cc(t_c)-\sin\delta_cc(t_c)\right]p_c(t_c)=1+\dfrac{m}{\delta_c}\dfrac{\partial f_c}{\partial m}-\dfrac{1}{\gamma L_o\delta_c^2}\dfrac{\partial f_c}{\partial m}p_c(t_c)\delta_cc(t_c)\cos\delta_cc(t_c).
\end{align}
In terms of $x_c(t_c)=\tan[\delta_cc(t_c)/2]>0$, it is simple to prove that, for the solutions that we have derived in Sec. \ref{sec:basics},
\begin{align}
p_c(t_c)&=\dfrac{1}{2}\gamma L_o\delta_c m \dfrac{1+x_c(t_c)^2}{x_c(t_c)},\\
\delta_cc(t_c)&=2\arctan x_c(t_c),\end{align}
and then 
\begin{align}
\cos\delta_cc(t_c)&=\dfrac{1-x_c(t_c)^2}{1+x_c(t_c)^2}.
\end{align}
Therefore, we can recast $F_{bc}$ as follows:
\begin{align}
F_{bc}(t_c)=1-\dfrac{m}{\delta_c}\dfrac{\partial f_c}{\partial m}\left[\arctan x_c(t_c)\dfrac{1-x_c(t_c)^2}{x_c(t_c)}-1\right].
\end{align}
The zeroes $t_c^{\rm ext}$ of $F_{bc}(t_c)$, i.e. the local extrema of $G_{c}(t_c)$, around which this function cannot be inverted univocally, then verify
\begin{align}
\arctan x_c^0\dfrac{1-(x_c^0)^2}{x_c^0}=1+\dfrac{\delta_c}{m}\left(\dfrac{\partial f_c}{\partial m}\right)^{-1},
\quad {\rm where} \quad   x_c^0=x_c(t_c^{\rm ext}) .\end{align}

Following the minimum area arguments of Refs. \cite{AOS,AOS2}, we will select the functions that appear in the definitions of the polymerization parameters in such a way that their on-shell values satisfy
\begin{align}
\delta_b=\left(\dfrac{\sqrt{\Delta}}{\sqrt{2\pi}\gamma^2m}\right)^{1/3}+o(m^{-1/3}),\quad \delta_c=\dfrac{1}{2L_o}\left(\dfrac{\gamma \Delta^2}{4\pi^2m}\right)^{1/3}+o(m^{-1/3}),\label{deltasm}
\end{align}
where $\Delta=4\sqrt{3}\pi G\gamma$ is the area gap in LQG and $o(\cdot\,)$ denotes terms that are subdominant with respect to the function in parentheses in the limit of large masses, $m\to \infty$. These subdominant terms appear because the expressions of the parameters derived in Refs. \cite{AOS,AOS2} involve a large mass expansion of certain minimum area conditions and, therefore, they are only valid for very massive black holes. 

\begin{figure}[h!]
\centering
\includegraphics[scale=0.6]{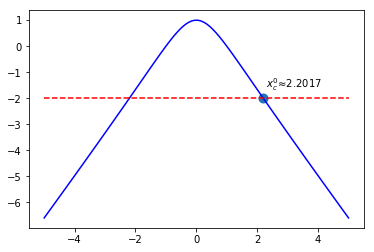}
\caption{Left and right hand sides of Eq. \eqref{x0} as functions of $x_c$. The continuous blue curve represents the left hand side, whereas the discontinuous red line is the dominant term for large masses on the right hand side. The intersection of both curves provides the points where the function $F_{bc}$ vanishes. Since $x_c^0>0$ by definition, only the intersection on the right semiaxis (i.e., the one marked with the approximate value of the corresponding $x_c^0$) is relevant for our discussion.}
\label{graph1}
\end{figure}

Employing the dependence of the on-shell parameters on $m$ shown above, we obtain that
\begin{align}
\arctan x_c^0\dfrac{1-(x_c^0)^2}{x_c^0}=-2+o(m^0).\label{x0}
\end{align}
Ignoring subdominant terms and recalling that $x_c^0$ is strictly positive, we see that there is just a single solution to this equation. In Fig. \ref{graph1}, we represent the functions on the left and right hand sides of the above expression, the intersection of which yields the only zero of $F_{bc}$. This zero is found to be $x_c^0\approx 2.2017$.

In conclusion, $G_c(t_c)$ is invertible at all times $t_c<0$ in the interior region except in a neighborhood of
\begin{align}
t_c^{\rm ext}=-\dfrac{1}{2}\ln \dfrac{8m[x_c^0+o(m^0)]}{\gamma L_o \delta_c}=t_c^\mathcal{T}-\dfrac{1}{2}\ln x_c^0+o\left(\ln\dfrac{\delta_c}{m}\right).\label{tcext}
\end{align}
This value of $t_c$ is reached beyond the transition surface (indeed, $t_c^{\rm ext}<t_c^{\mathcal{T}}$ because $x_c^0>1$). We also note that the numerical value of $t_c^{\rm ext}$ receives subdominant corrections coming from higher-order terms in the on-shell expressions of the polymerization parameters, as shown by the inclusion of the term $o[\ln(\delta_c/m)]$ in the previous equation. As a consequence of these invertibility properties, $t_c$ can be written in terms of $t_b$ as
\begin{align}
t_c=G_c^{-1}[G_b(t_b)],
\end{align}
as long as we are away from $t_c^{\rm ext}$. Around this value of the angular time, it might even be the case that the effective spacetime metric could be acceptably defined, e.g. if $G_b$ can be inverted there, so that we can express the radial time in terms of the angular time instead. A single time description of the spacetime geometry would be attainable if we could satisfactorily combine local inversions covering all of the interior region. We note, however, that this imposes severe restrictions on the images of $G_b$ and $G_c$. For the moment, we will study the zeroes of $F_{cb}$ and show that in fact none of them corresponds to $t_c^{\rm ext}$ with our definitions. As we have already pointed out, the discussion of these zeroes will be also extremely helpful to describe the qualitative behavior of the function $G_b$ in the interior region and, hence, of its image.   

\begin{figure}[h!]
\centering
\includegraphics[scale=0.6]{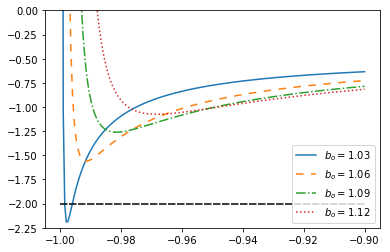}
\caption{Left and right hand sides of Eq. \eqref{xb} as functions of $x_b^0$. The various curves correspond to the left hand side of Eq. \eqref{xb} for different values of $b_o$, while the discontinuous black line is the value of the right hand side of the equation.} 
\label{graph2}
\end{figure}

According to the definitions \eqref{Fijonshell} and \eqref{dObddeltab}, the function $F_{cb}(t_b)$ can be written as
\begin{align}
F_{cb}(t_b)=1+\dfrac{1}{2\gamma L_o\delta_b^2}\dfrac{\partial f_b}{\partial m}\left[1-\dfrac{\gamma^2\delta_b^2}{\sin^2\delta_b b(t_b)}\right]\left[\delta_bb(t_b)\cos\delta_bb(t_b)-\sin\delta_bb(t_b)\right]p_b(t_b).
\end{align}
The expression of the radial partial Hamiltonian \eqref{Ob} implies that
\begin{align}
\dfrac{\gamma^2\delta_b}{\sin\delta_bb(t_b)}p_b(t_b)=-2\gamma L_o m-\dfrac{\sin\delta_bb(t_b)}{\delta_b}p_b(t_b).
\end{align}
Therefore,
\begin{align}
F_{cb}(t_b)=1-\dfrac{m}{\delta_b}\dfrac{\partial f_b}{\partial m}+\dfrac{1}{2\gamma L_o \delta_b}\dfrac{\partial f_b}{\partial m}\left\{\left[1-\dfrac{\gamma^2\delta_b^2}{\sin^2\delta_bb(t_b)}\right]\dfrac{\delta_bb(t_b)\cos\delta_bb(t_b)}{\delta_b}p_b(t_b)-2\dfrac{\sin\delta_bb(t_b)}{\delta_b}p_b(t_b)\right\}.\label{caminoFcb}
\end{align}
Using the dynamical solutions presented in Sec. \ref{sec:basics}, $\cos\delta_bb(t_b)$ can be recast as
\begin{align}
\cos\delta_bb(t_b)=\dfrac{1+b_o x_b(t_b)}{1+b_o^{-1}x_b(t_b)},\label{cosdeltabbxb}
\end{align}
where $-1<x_b(t_b)=\tanh(b_ot_b/2)\leq 0$. Notice that Eq. \eqref{cosdeltabbxb} reveals that $-b_o<\cos\delta_bb(t_b)\leq 1$, with a lower bound smaller than the usual one on the cosine function of a real variable. This difference points towards the fact that the radial canonical variables become imaginary at some point along the evolution. This phenomenon is intimately related to the existence of a horizon beyond the transition surface in the original model \cite{AOS,AOS2}. We will concentrate our analysis on values of the radial time for which the radial connection variable $b$ is real and, thus, its trigonometric functions are bounded in the standard way. These values are those corresponding to $x_b(t_b)\in[-2b_o/(1+b_o^2),0]$, to $\delta_bb(t_b)\in[0,\pi]$ or, equivalently, to
\begin{align}
t_b\in\left[t_{b}^{\rm WH},0\right],\quad t_b^{\rm WH}=-\dfrac{4}{b_o}\arctanh \dfrac{1}{b_o},\label{WH}
\end{align}
where $t_{b}^{\rm WH}$ denotes the position of a white hole horizon according to the interpretation of Refs. \cite{AOS,AOS2}. From now on, we will restrict our discussion to the \emph{genuine interior region}, comprised between the sections identified as black and white horizons. Taking into account this restriction and Eq. \eqref{pbeff}, we have that
\begin{align}
\delta_bb(t_b)&=\arccos\left[\dfrac{1+b_ox_b(t_b)}{1+b_o^{-1}x_b(t_b)}\right],\\
\sin\delta_bb(t_b)&=\gamma\delta_b\dfrac{\sqrt{-x_b(t_b)[2b_o+(1+b_o^2)x_b(t_b)]}}{b_o+x_b(t_b)},\\
p_b(t_b)&=-2L_o m\dfrac{[b_o+x_b(t_b)]\sqrt{-x_b(t_b)[2b_o+(1+b_o^2)x_b(t_b)]}}{b_o^2[1-x_b(t_b)^2]}.
\end{align}
Notice that, in fact, both $p_b$ and $\sin\delta_b b$ vanish at the white horizon. Introducing these expressions in Eq. \eqref{caminoFcb}, we get
\begin{align}
F_{cb}=&1-\dfrac{2m}{\delta_b}\dfrac{\partial f_b}{\partial m}\left\{\dfrac{1}{2}+\dfrac{1}{2\sqrt{b_o^2-1}}\left[1+\dfrac{(b_o+x_b)^2}{2b_o x_b+(1+b_o^2)x_b^2}\right]\arccos\left(\dfrac{1+b_ox_b}{1+b_o^{-1}x_b}\right)\dfrac{1+b_ox_b}{b_o(1-x_b^2)}\sqrt{-x_b[2b_o+(1+b_o^2)x_b]}\right.\nonumber\\
&\left.+\dfrac{2b_ox_b+(1+b_o^2)x_b^2}{b_o^2(1-x_b^2)}\right\}.
\end{align}
Then, any zero $t_b^{\rm ext}$ of $F_{cb}$ must satisfy that, with $x_b^0=x_b(t_b^{\rm ext})$ (and up to subdominant corrective terms),
\begin{align}
\dfrac{1}{2\sqrt{b_o^2-1}}\left[1+\dfrac{(b_o+x_b^0)^2}{2b_o x_b^0+(1+b_o^2)(x_b^0)^2}\right]\arccos\left(\dfrac{1+b_ox_b^0}{1+b_o^{-1}x_b^0}\right)&\dfrac{1+b_ox_b^0}{b_o[1-(x_b^0)^2]}\sqrt{-x_b^0[2b_o+(1+b_o^2)x_b^0]}\nonumber\\
&+\dfrac{2b_ox_b^0+(1+b_o^2)(x_b^0)^2}{b_o^2[1-(x_b^0)^2]}=-2,\label{xb}
\end{align}
where we have used the on-shell dependence on $m$ of the radial polymerization parameter. In Fig. \ref{graph2}, we represent the left and right hand sides of Eq. \eqref{xb}. Their intersections provide the (image under $x_b$ of the) zeroes of the function $F_{cb}$.

By inspecting Fig. \ref{graph2}, we observe the following. As $b_o$ approaches one (i.e. for small values of $\delta_b$ or, equivalently, large masses $m$), the left hand side of Eq. \eqref{xb} displays a minimum that is more pronounced and more displaced toward negative values of $x_b^0$, and thus of $t_b$. As a result, we find that $F_{cb}$ does not have any zero for masses $m$ below a critical threshold, value at which the minimum of the left hand side of the studied equation is pronounced enough to barely lead to an intersection. Numerically, we observe that this threshold value corresponds to $b_o^{\rm crit}\approx 1.032$ (or, equivalently, to a value of $m$ of approximately 13 Planck masses). Since the model is specifically adapted to describe black holes that are very massive, in these cases there will always exist two zeroes $x_{b,0}^{(1)}$ and $x_{b,0}^{(2)}$, such that $x_{b,0}^{(1)}>x_{b,0}^{(2)}$. Thus, the primitive $G_b$ is locally invertible around all radial times $t_b<0$ except at two instants, $t_{b,\rm ext}^{(1)}$ and $t_{b,\rm ext}^{(2)}$.  

Let us verify now whether any of these points corresponds to the value of the angular time around which $G_c$ cannot be inverted for any value of the mass $m$.  We recall that the correspondence between the times $t_b$ and $t_c$ is dictated by the relation $G_b(t_b)=G_c(t_c)$. Therefore, we have to check whether or not $G_c(t_c^{\rm ext})$ differs from $G_{b}(t_{b,\rm ext}^{(1)})$ and $G_b(t_{b,\rm ext}^{(2)})$, for any (large) value of the mass. However, since neither the zeroes of $F_{cb}(t_b)$ nor $G_{b}(t_b)$ itself can be written in terms of elementary functions, we will have to resort to a seminumerical argument.

In the angular sector, it is possible to integrate $F_{bc}(t_c)$ explicitly. Indeed, up to subdominant terms,
\begin{align}
G_c(t_c)&=-t_c+\dfrac{1}{3\gamma L_o\delta_c m}\int_{t_c}^0dt_c'[\delta_cc(t_c')\cos\delta_cc(t_c')-\sin\delta_cc(t_c')]p_c(t_c')\nonumber \\
&=-t_c-\dfrac{1}{6}\int_{x_c(t_c)}^{x_c(0)}dx_c\left[\left(\dfrac{1}{x_c^2}-1\right)\arctan x_c-\dfrac{1}{x_c}\right].
\end{align}
Integrating by parts, we obtain the following expression:
\begin{align}
G_c(t_c)&=-t_c-\dfrac{1}{6}\left\{\left[x_c(t_c)+\dfrac{1}{x_c(t_c)}\right]\arctan\left[x_c(t_c)\right]-\left[x_c(0)+\dfrac{1}{x_c(0)}\right]\arctan\left[x_c(0)\right]\right\}.\label{Gcexplicit}
\end{align}
In particular, the value of the primitive $G_{c}(t_c)$ at its only extremum is obtained by replacing $x_c(t_c)$ with $x_c^0$, and
recalling that $x_c(0)=\gamma L_o \delta_c/(8m)$. Like $x_c(0)$, this value $G_{c}(t_c^{\rm ext})$ depends on the black hole mass $m$, both directly and indirectly through its dependence on $\delta_c$.

\begin{figure}[h!]
\centering
\includegraphics[scale=0.65]{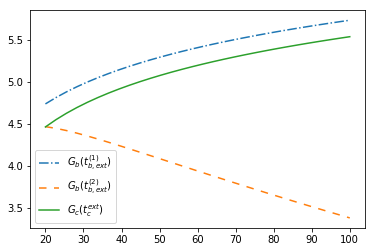}
\caption{Primitives evaluated at the zeroes of their derivatives for values of the mass $m$, represented on the horizontal axis, ranging between 20 and 100 Planck masses. $G_c(t_c^{\rm{ext}})$, $G_b(t_{b,\rm{ext}}^{(1)})$, and $G_{b}(t_{b,\rm{ext}}^{(2)})$ are depicted by means of a continuous green curve, a dash-dotted blue curve, and a dashed orange curve, respectively.} 
\label{graph3}
\end{figure}

It would be desirable to have a similar expression for the evaluation of the radial primitive $G_b$ at its two extrema. Nevertheless, as we mentioned before, it does not even seem possible to write $G_b$ in terms of elementary functions, although its existence is guaranteed by the integrability argument discussed at the beginning of this section. For the sake of completeness, the expression of $G_b(t_b)$ is explicitly given by
\begin{align}
G_b(t_b)&=-t_b-\dfrac{1}{6\gamma L_o\delta_b m}\int_{t_b}^0dt'_b\left[1-\dfrac{\gamma^2\delta_b^2}{\sin^2\delta_bb(t_b')}\right]\left[\delta_bb(t_b')\cos\delta_bb(t_b')-\sin\delta_bb(t_b')\right]p_b(t_b')\nonumber\\
&=-t_b+\dfrac{2}{3\gamma \delta_b }\int_{x_b(t_b)}^{0}dx_b'\left[1+\dfrac{(b_o+x_b')^2}{2b_ox'_b+(1+b_o^2){x'_b}^2}\right] \dfrac{(b_o+x'_b)\sqrt{|2b_ox'_b+(1+b_o^2){x'_b}^2|}}{b_o^3(1-{x'_b}^2)^2}\nonumber\\
&\times\left[\arccos\left(\dfrac{1+b_ox'_b}{1+b_o^{-1}x'_b}\right)\dfrac{1+b_ox'_b}{1+b_o^{-1}x'_b}-\gamma\delta_b\dfrac{\sqrt{|2b_ox'_b+(1+b_o^2){x'_b}^2|}}{b_o+x'_b}\right],\label{Gbcuadratura}
\end{align}
where subdominant terms have been omitted. If we were to reintroduce them, these terms [sourced by the neglected corrections in Eq. \eqref{deltasm}] would modify the factor that multiplies the integral in the previous equation, as well as the parameters $b_o$ that appear inside the integral. Let $G_b(t_{b,\rm ext}^{(a)})$ be the evaluation of Eq. \eqref{Gbcuadratura}  at the zeroes of $F_{cb}(t_b)$, $t_{b,\rm ext}^{(a)}$ with $a=1,2$. As in the case of $G_c(t_c^{\rm ext})$, the value $G_b(t_{b,\rm ext}^{(a)})$ depends on the mass $m$, this time through the mass dependence of $\delta_b$ and of the zeroes of $F_{cb}$ (recall that $b_o$ enters the equation satisfied by $x_{b,0}^{(a)}$). Thus, for each value of the mass, we need to numerically solve Eq. \eqref{xb} and then evaluate $G_b$ at its solutions, evaluation which also has to be done by numerical methods. This procedure allows us to represent the two curves $G_b(t_{b,\rm ext}^{(a)})$ as functions of $m$. Their intersections with $G_c(t_c^{\rm ext})$ indicate potential obstructions to our formalism. The result of this numerical computation can be seen in Fig. \ref{graph3}.

The plot in Fig. \ref{graph3} shows that $G_b(t_{b,\rm ext}^{(1)})$ is larger than $G_{c}(t_c^{\rm ext})$ in the entire interval of studied masses. The only possible intersection is, then, between $G_c(t_{c}^{\rm ext})$ and $G_b(t_{b,\rm ext}^{(2)})$, which indeed is found to occur for a value of the mass $m$ around 20 Planck masses. Beyond this value, the curves of the primitives $G_i$ evaluated at their extrema do not intersect one another, at least for the considered masses. Moreover, the behavior displayed by these curves suggests that this conclusion can be expected to apply to larger values of $m$ as well.

Therefore, the results of this section guarantee that, at every point of the interior region, it is possible to invert locally at least one of the time functions $G_b$ or $G_c$ (in terms of their respective arguments) provided that the mass of the black hole under consideration is larger than some 20 Planck masses, which includes the whole regime of validity of the original model put forward in Refs. \cite{AOS,AOS2,AO}. However, as we have already commented, these integrability and local invertibility conditions are necessary but still not sufficient to guarantee a single time formulation in the whole of the studied region. At this stage of the discussion, we can only conclude that a single time reformulation is possible in the region covered by the time intervals associated with the intersection of the images of $G_b$ and $G_c$ (as long as we focus on sufficiently large masses $m$). Determining whether or not this region coincides with the whole interior is the aim of Sec. \ref{subsec:imageG}.

\subsection{Images of the two time functions $G_i$} \label{subsec:imageG}

We already know that $G_b(t_b)$ and $G_c(t_c)$ exist and that, at every time corresponding to the interior region, at least one of them is invertible if $m$ is sufficiently large. Nevertheless, this does not mean yet that the effective metric can be defined properly in a single time formalism. In particular, the results about the local invertibility of $G_b$ and $G_c$ are still insufficient to express one of the two times, the radial or the angular one, in terms of the other if the images of the interior region under $G_b$ and $G_c$ do not match. The implicit relation $G_b(t_b)=G_c(t_c)$ can only be satisfied for values of the times that correspond to the intersection of the images of these two functions. To complete our discussion about the invertibility of the relation between the radial and angular times, in this section we study the images of the functions $G_i$. For this, the results about the extrema of $G_i$ obtained in the previous subsection will be of the greatest help.

Let us start by considering the angular primitive $G_c$, of which we have an explicit expression. According to Eq. \eqref{Gcexplicit}, it is immediate to see that $G_c(0)=0$ and $\lim_{t_c\to -\infty}G_c=-\infty$. Furthermore, for small values of $t_c$ (i.e. close to the black horizon), the function $G_c$ is positive and behaves in an approximately linear fashion. Nonetheless, for sufficiently negative values of $t_c$, the behavior becomes that of a decreasing exponential. This, together with the fact that $G_c$ is continuous, implies that this function reaches an odd number of local extrema. In fact, we already know that this function has just a single extremum: a maximum  $G_{c,\rm max}=G_c(t_c^{\rm ext})$ at a time $t_c^{\rm ext}$ for which $x_c(t_c^{\rm ext})\approx 2.2017$ [see Eqs. \eqref{tcext} and \eqref{Gcexplicit}]. Subsequently, the image under the angular primitive of the negative real semiaxis $t_c\leq 0$, which contains the interior region, is 
\begin{align}
\mathcal{R}(G_c)=(-\infty,G_{c,\rm max}].\label{RGc}
\end{align}
Recall that a potential white horizon is found for a finite value of $t_b$ and, given the continuity properties of $G_b$ and $G_c$, this boundary will also correspond to a finite value of $t_c$. As a consequence, the image of $G_c$ restricted to the interior region, $\mathcal{R}(G_c)|_{\rm int}$, must be a bounded interval and therefore have a finite lower end.
 
From expressions \eqref{Gcexplicit} and \eqref{tcext} for $G_c$ and the time $t_c^{\rm ext}$, we can understand how the position of the maximum is affected by a variation of the mass of the black hole under consideration. Indeed, as revealed by the signs of the derivatives of $G_c$ and of $t_c^{\rm ext}$ with respect to $m$, the maximum gets displaced towards more negative values of $t_c$ as the mass increases, whereas the height of the maximum grows. In what follows, we will carry out a similar analysis of $G_b$ to understand its behavior as a function of $t_b$ and how it is modified by a change in the value of $m$.

With our definition of $G_b$ in Eq. \eqref{Gbcuadratura}, this function vanishes at the black horizon. Additionally, it also turns out to be positive and essentially linear in a neighborhood of this boundary, as we commented that it is the case for its angular counterpart $G_c$. Its asymptotic form, however, differs greatly from that of $G_c$. Instead of decreasing exponentially, it can be seen that it tends to positive infinity. As a result, the image of the negative real semiaxis under $G_b$ will certainly contain the positive, real half-line, $\mathcal{R}(G_b)\supseteq [0,\infty)$. If we restrict our attention to the interior region (that is, no farther than the white horizon, located at a finite value of $t_b$), this conclusion is modified. The correct statement in that case would be that $\mathcal{R}(G_b)|_{\rm int}\supseteq [0,G_{b, \rm sup}]$, where $G_{b,\rm sup}$ denotes the supremum of $G_b$ in the region under consideration, which could correspond to either its value at one of its extrema, $G_b(t_{b,\rm ext}^{(a)})$, or its value at the assumed white horizon, $G_b(t_b^{\rm WH})$. In any of these cases, $G_{b,\rm sup}$ turns out to be finite, because $G_b$ is continuous. 

\begin{figure}[h!]
\centering
\includegraphics[scale=0.65]{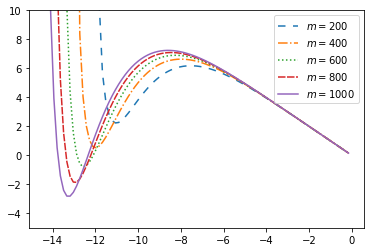}
\caption{$G_b$ as a function of the radial time $t_b$ for different values of the mass parameter $m$. We have taken $G=1$, $L_o=1$, and the standard values of $\gamma$ and $\Delta$.}
\label{graph5}
\end{figure}

Moreover, as we have already seen, $G_b$ exhibits two extrema\footnote{Numerical results suggest that the expected white horizon lies always beyond the minimum, which means that both extrema belong to the interior region, at least for masses within the studied range.} at the solutions of Eq. \eqref{xb}. The only possibility that is compatible with the continuity of $G_b$ and its asymptotic behavior is that the extremum that is closer to the black horizon is a maximum and the one farther away from it, a minimum. Their behavior as $m$ increases is displayed in Fig. \ref{graph5}. 

By inspecting this graph, we conclude that the difference in height between the maximum and the minimum increases as the mass gets larger, and that both extrema shift towards more negative values of $t_b$ as $m$ grows. In addition, we observe that the value of the minimum, $G_{b, \rm min}$, becomes negative once the mass surpasses a critical value (numerically found to be around 460 Planck masses). Hence, for values of the mass beyond this threshold, the image of the radial primitive would be of the form
\begin{align}
\mathcal{R}(G_b)|_{\rm int}=[G_{b,\rm min},G_{b,\rm sup}].\label{RGb}
\end{align}

Obviously, the continuity of $G_b$ ensures that $G_{b, \rm min}$ is always finite, for any given finite value of $m$. In agreement with our comments above, the relation between the radial and angular times then implies that $G_c(t_c)$ must also be bounded from below in the interior region, the bound being precisely $G_{b,\rm min}$ in the sector of large masses $m$ of interest\footnote{More generally, $\mathcal{R}(G_c)|_{\rm int}=[G_{b,\rm inf},G_{c,\rm max}]$, where the infimum of the radial primitive $G_{b,\rm inf}$ is either zero or $G_{b,\rm min}$, depending on whether $m$ is above a certain mass threshold or not.}. We emphasize that the reason for this lower bound in the interior is that the characterization of the white horizon is based on the vanishing of $b$ and $p_b$ (just as the black horizon), so that its location is naturally determined in terms of the radial time [see Eq. \eqref{WH}]. With this picture in mind and the considered relation between times, we see that, for sufficiently massive black holes, i) $t_c$ decreases monotonically until it reaches a value $G_c^{-1}[G_b(t_{b, \rm ext}^{(2)})]$, where $G_{c}$ attains its minimum\footnote{It is straightforward to see that this minimum, where $dt_c/dt_b$ vanishes, signals the presence of a local extremum in the area of the coordinate 2-spheres, $4\pi p_c(t_b)$.}; and ii) $t_c$ then starts to increase until it reaches the value $G_c^{-1}[G_{b}(t_b^{\rm WH})]$, corresponding to the end of the interior region. Therefore, we conclude that, for sufficiently large masses $m$, 
\begin{align}
\mathcal{R}(G_c)|_{\rm int}=[G_{b,\rm min},G_{c,\rm max}],\quad \mathcal{R}(G_b)|_{\rm int}=[G_{b,\rm min},G_{b,\rm sup}], \label{RGcRGb}
\end{align}
where $G_{b,\rm sup}=\textrm{max} [G_{b}(t_b^{\rm WH}),G_{b,\rm max}]$. As an immediate consequence of the previous expressions, any possible discrepancies in the images of $G_b$ and $G_c$ must be located at their upper endpoints. 

\begin{figure}[h!]
\centering
\includegraphics[scale=0.65]{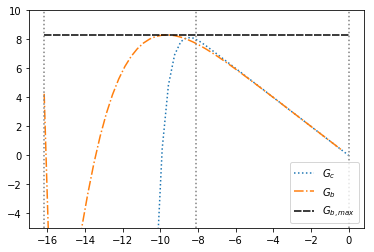}
\caption{$G_b$ and $G_c$ as functions of their respective times $t_b$ and $t_c$ (identified with the variable of the horizontal axis) for $m=5000$. The dotted gray lines denote, from left to right, the position of the potential white horizon, of the transition surface, and of the black horizon in the time variable in terms of which they are naturally defined (i.e., in radial time in the case of the horizons and in angular time in the case of the transition surface). 
We have taken $G=1$, $L_o=1$, and the standard values of $\gamma$ and $\Delta$.}
\label{graph6}
\end{figure}

Once the behavior of $G_b$ and $G_c$ has been understood, let us conclude our discussion about how their images compare with each other. Representing the two primitive functions at the same time for a value of $m$ much larger than the Planck mass, namely $m=5000$, yields the result displayed in Fig. \ref{graph6}. As we can see in that figure, both $G_b$ and $G_c$ are essentially equal until the transition surface, where quantum effects start to become relevant. So, $t_b\approx t_c$ until that surface. The closer we get to the black horizon, the better this approximation becomes. In this subregion, we can achieve a partial reconciliation of our formalism with the results of the original model given that, in this regime, our dynamical solutions are identical to the ones in Refs. \cite{AOS,AOS2}. Nevertheless, the single time coordinate in which they are written (i.e. the approximately coincident value of $t_b$ and $t_c$) is not the one considered by the authors of that reference, as manifest by the presence of the factor $1/C_{ij}^2$ in the time component of the effective metric, factor that we will study in the next section.

In contrast with the above comments, $G_b$ and $G_c$ differ wildly beyond the transition surface. In particular, as we have shown, the radial primitive displays a second extremum in that subregion, which leads to a non-monotonic relation between the two times, as we commented in the paragraph below Eq. \eqref{RGb}. Moreover, we observe that $G_{b,\rm sup}>G_{c,\rm max}$. It is worth emphasizing that this conclusion has been found to hold in the entire interval of values of the mass under consideration. Indeed, the value of $G_b$ at its maximum (i.e. at $t_{b,\rm ext}^{(1)}$), which corresponds to $G_{b, \rm sup}$ for very massive black holes according to our numerical analysis, turns out to be always larger than $G_{c,\rm max}$ for the values of $m$ that we have been considering (see Fig. \ref{graph3}). Furthermore, given the behavior exhibited by the primitive functions in the studied interval of values of $m$, it can be expected that $G_{b,\rm sup}$ remains larger than $G_{c,\rm max}$ for more massive black holes as well.

It is now straightforward to see that the images of the interior region under the two functions $G_c$ and $G_b$, as we have defined them in Eq. \eqref{RGcRGb}, do not coincide for any finite value of $m$ in the studied interval of masses. Indeed, the fact that $G_{b,\rm sup}$ is by definition larger than or equal to the maximum of $G_b$, and that this latter quantity exceeds $G_{c,\rm max}$, means that there exist values of the radial time with a positive image that find no match in terms of angular times. Thus, a satisfactory inversion of the considered time relation in the whole interior region is not possible.

In conclusion, despite the integrability and good local invertibility properties of the functions $F_{ij}$ and $G_i$, respectively, the fact that the intersection of the images of $G_i$ does not cover the totality of the interior region implies \emph{an obstruction} to the formalism derived from our proposal to define the polymerization parameters \eqref{deltaiObOc}. Indeed, our formalism cannot be extended to encompass the whole interior region, at least in its present form, in the sense that it fails to provide an effective spacetime metric that is well defined everywhere by virtue of the existence of a subregion where a satisfactory relation between the radial and angular times cannot be established. At this stage, however, we recall that there exists a freedom in the formalism which we have not yet exploited, namely, the freedom to select the origins of the two times independently. We will see in Sec. \ref{sec:redefinitionorigins} that we can actually make  the images of $G_b$ and $G_c$ match with an appropriate choice of these origins, removing in this way the obstruction that we have found to define an effective geometry in the totality of the interior.

\subsection{The time component of the metric} \label{subsec:analysisC}

For the sake of completeness in our analysis, we want to study in this section the behavior of the factor $C_{ij}$ along dynamical trajectories, with the aim of showing that this factor does not introduce additional obstructions to our two-time formalism. We know from the expressions in Eq. \eqref{Cij} that the factors $C_{ij}$ exclusively depend on the derivatives of the partial Hamiltonians with respect to their associated polymerization parameter and on the derivatives of these parameters with respect to the other Hamiltonian. More concretely, these factors can be written as follows:
\begin{align}
C_{ij}=\dfrac{F_{ij}}{\Xi},\qquad \Xi=(1-\Delta_{bb})(1-\Delta_{cc})-\Delta_{bc}\Delta_{cb}.\label{Xi}
\end{align}
Given this structure (and the continuity of all the involved functions), there seems to be three situations that need to be examined carefully in order to rule out the possibility that the resulting effective metric is divergent or degenerate. Firstly, we have to consider the points where $F_{ij}$ vanishes, corresponding to (possibly apparent) divergences in the metric. Secondly, it is also important to study the zeroes of the denominator $\Xi$, associated with points where the effective metric becomes degenerate (at least in the adopted coordinates). Finally, we should contemplate the possibility that $F_{ij}$ and $\Xi$ vanish at the same time, which would lead to an indeterminate situation.

As we have already seen, the points where the function $F_{ij}$ vanishes are those in which $G_j$ (i.e. the primitive function associated with $-F_{ij}$) cannot be inverted locally. This, together with the fact that (a power of) $F_{ij}$ enters the effective metric only when it is written in terms of the time $t_i$, implies that this situation does not constitute a source of problems. For the sake of clarity, let us focus our discussion on one particular case to show why. Let us ask whether the effective metric written in terms of the radial time $t_b$ is well defined:
\begin{align}
ds^2=-\dfrac{\gamma^2\delta_b^2p_c[t_c(t_b)]}{\sin^2\delta_bb(t_b)}\dfrac{\Xi[t_b,t_c(t_b)]^2}{F_{bc}[t_c(t_b)]^2}dt_b^2+\dfrac{p_b^2(t_b)}{L_o^2p_c[t_c(t_b)]}dx^2+p_c[t_c(t_b)]d\Omega^2,
\end{align}
where $t_c(t_b)$ should be understood as $t_c(t_b)=G_c^{-1}[G_b(t_b)]$. It is crucial to keep in mind that the effective metric can only be recast in this form \emph{away} from the zeroes of $F_{bc}$, where $G_c$ is invertible and the angular time can be expressed in terms of $t_b$. Thus, it is obvious that $F_{bc}$ never vanishes in the region that can be covered with an effective metric written in terms of the radial time. The same argument applies to the case where the effective metric is written in terms of $t_c$. In that case, the potentially problematic factor that appears in the metric is (a power of) $F_{cb}$, but that function is ensured to be non-vanishing since it is a necessary condition to express the effective metric in terms of the angular time. 

Notice that the fact that the behavior of $F_{ij}$ is harmless rules out both the first and the third potential sources of pathologies mentioned in the paragraph below Eq. \eqref{Xi}. The numerator of $C_{ij}$ does not vanish in the region where the time component of the metric can be written in an appropriate way as to make this function appear in it. Therefore, the situation where both the numerator and the denominator vanish at the same time cannot occur either. Thus, the only potential situation that may prevent a well-behaved factor $1/C_{ij}^2$ is the possibility that $\Xi$ becomes zero at some point. We recall that the value of $\Xi$ has been, in fact, assumed to be non-zero in the derivation of the equations of motion. However, once the time redefinitions are introduced and the resulting dynamical equations are solved, the dynamical solutions are well defined even if $\Xi=0$. Hence, we can employ a continuity argument to extend our formalism so that it includes the case where this restriction is absent and then study the resulting behavior of $\Xi$ along any possible dynamical trajectory.

The points at which the denominator $\Xi$ vanishes are determined by the equation
\begin{align}
\left(1-\dfrac{\partial f_b}{\partial O_b}\dfrac{\partial O_b}{\partial \delta_b}\right)\left(1-\dfrac{\partial f_c}{\partial O_c}\dfrac{\partial O_c}{\partial \delta_c}\right)=\dfrac{\partial f_b}{\partial O_c}\dfrac{\partial O_b}{\partial \delta_b}\dfrac{\partial f_c}{\partial O_b}\dfrac{\partial O_c}{\partial \delta_c}.\label{63}
\end{align}
Recall that, after evaluating the polymerization parameters on shell and assuming that they are at least of class $\mathcal{C}^1$,
\begin{align}
\dfrac{\partial f_i}{\partial O_b}+\dfrac{\partial f_i}{\partial O_c}=\dfrac{\partial f_i}{\partial m}.
\end{align}
Then, we can exploit the off-shell freedom of our formalism to choose freely one of the derivatives on the left hand side of the previous equation. Once that choice has been made, the other derivative is immediately fixed so that the sum of both is equal to the derivative of the parameters \eqref{deltasm} with respect to the mass. Expressing in this way the non-diagonal derivatives $\partial f_i/\partial O_j$, we can rewrite Eq. \eqref{63} as
\begin{align}
\left(F_{bc}\dfrac{\partial O_b}{\partial \delta_b}\right) \dfrac{\partial f_b}{\partial O_b}+\left(F_{cb}\dfrac{\partial O_c}{\partial \delta_c}\right) \dfrac{\partial f_c}{\partial O_c}+\left(F_{bc}+F_{cb}-F_{bc}F_{cb}\right)=0,\label{convenientform}
\end{align}
where the objects in parentheses can be understood as the coefficients (in general, dependent on the mass $m$ and on the phase space point) of an equation linear in $\partial f_b/\partial O_b$ and $\partial f_c/\partial O_c$. On the one hand, it can be immediately seen that $F_{bc}=F_{cb}=0$ provides a trivial solution. However, as discussed in Sec. \ref{subsec:analysisC}, $F_{bc}$ and $F_{cb}$ do not vanish simultaneously in the regime of interest of the model, where $m$ is very large compared to the Planck mass. On the other hand, it is clear that, as long as the first two coefficients do not vanish at the same time, it is possible to exploit the freedom of our formalism to select $\partial f_i/\partial O_i$ in such a way that the above equation is not satisfied (in other words, in such a way that $\Xi\neq 0$). It is straightforward to prove that the first two coefficients of Eq. \eqref{convenientform} cannot be zero at the same point along any dynamical trajectory or at least not in a harmful way. In the first place, we know that the two functions $F_{ij}$ cannot vanish concurrently. Therefore, only two options are available for these coefficients to be zero: either A) one of the pairs $(F_{ij},\partial O_j/\partial \delta_j)$ vanishes at the same point or B) both derivatives $\partial O_i/\partial \delta_i$ do. The first possibility is immediately ruled out because, if $\partial O_j/\partial \delta_j$ is zero, then $F_{ij}=1$ by definition. While the second possibility does happen (namely in a neighborhood of the black horizon, where the primitives $G_i$ exhibit an approximately linear behavior), in that case both functions $F_{ij}$ would be equal to one and we can immediately realize that Eq. \eqref{convenientform} would not hold. Indeed, in that situation, the left hand side of Eq. \eqref{convenientform} would be given by $F_{bc}+F_{cb}-F_{bc}F_{cb}$, which would be equal to one and, thus, would certainly not vanish. In conclusion, we can always select the two off-shell derivatives $\partial f_i/\partial O_i$ in such a way that the denominator $\Xi$ is different from zero along any given dynamical trajectory.

The results derived in this section ensure that there is enough freedom in our formalism to guarantee that, if the factor $1/C_{ij}^2$ appears in the purely time component of the effective metric, this factor is finite and non-vanishing. This is a fundamental difference with respect to the approach proposed in Ref. \cite{N}, the ideas of which our proposal generalizes. In the approach of that reference, the absence of cross-derivatives of the polymerization parameters makes inevitable that one hits a singularity in the time evolution. In conclusion, the only identified obstruction to the present form of our two-time formalism does not affect the time component of the metric, but is rather due to problems in obtaining a global inversion of the relation between the radial and the angular times.

\section{Change of origin of the angular time} \label{sec:redefinitionorigins}

In the previous section, we argued that it is apparently impossible to extend our formalism to the whole interior region in a satisfactory way owing to the difference between the images of the two time functions $G_b$ and $G_c$. One could still wonder whether this problem could be circumvented by making use of the freedom to fix independently the origins of the two times. We note that the integration constants of the dynamical solutions were chosen in Refs. \cite{AOS,AOS2} in such a way that the radial variables $b$ and $p_b$ vanish at $t_b=0$, allowing an interpretation of that surface as a black horizon. Since we have followed their prescription for the selection of integration constants, we have also fixed the horizon at $t_b=0$. Nevertheless, it is worth remarking that this argument does not involve the angular time. So, we do not need to identify $t_c=0$ as well with the black horizon, as we have actually done until now. In general relativity, the linearity of the relation between the radial and the angular times makes irrelevant a constant shift of one of the times with respect to the other and, for simplicity, this shift is set equal to zero. But the non-linearity of the time relation in our effective model changes the situation. In the present section we will consider the possibility of a different choice of origin for the angular time and discuss if this can help to solve the obstruction that we have found for our formalism.

Let us recall that the definitions of the radial and angular times \eqref{defti} only involve their differentials. A different choice of origin for the angular time only entails a trivial displacement at the level of the angular part of the dynamical solutions, but this turns out to have an interesting effect on the implicit relation between the radial and angular times, as we have anticipated. Indeed, a change of origin for the angular time implies a modification of the integration limit in the definition of the primitive function $G_c$. As a result, the implicit relation $G_b(t_b)=G_c(t_c)$ is modified to
\begin{align}\label{gmdef}
G_b(t_b)=G_c(t_c)+G_m,
\end{align}
where $G_m=\int_0^{t_c^0}dt_c'F_{bc}(t_c')$ is the constant contribution of the interval of angular times between the old and the new origins. 

Let us now show that we can make $G_m$ compensate the difference between $G_{c,\rm max}$ and $G_{b,\rm sup}$ and, in this manner, ensure that the images of these two functions have the same upper endpoint and, therefore, coincide. It is worth noting that, for such a redefinition, $G_m$ needs to be chosen differently for each value of $m$. In the sector of very massive black holes, the supremum of $G_b$ is just its value at its maximum (see, for instance, Fig. \ref{graph6}). Hence, taking into account our considerations above, we can adopt a (mass-dependent) redefinition of the origin of the angular time such that
\begin{align}\label{gmiden}
G_c(t_c^{\rm ext})+ G_m =G_b(t_{b,\rm ext}^{(1)}).
\end{align}
Indeed, our numerical results support that this can be achieved for all masses in the sector of interest. On the one hand, one observes that the difference $G_b(t_{b,\rm ext}^{(1)})-G_c(t_c^{\rm ext})$, which is positive for large masses, does not increase with the mass $m$. On the other hand, by extending $G_c(t_c)$ to positive values of the angular time and using the integral definition of $G_m$ given below Eq. \eqref{gmdef}, one can see that this quantity can at least take all positive values up to $G_{c,\rm max}$, which is sufficient to guarantee that the matching imposed in Eq. \eqref{gmiden} can be satisfied (see e.g. Fig. \ref{graph3}). 

After this readjustment, we have that the implicit relation between times now holds in a neighborhood of the coincident maxima of the functions $G_b$ and $G_c$, where $F_{bc}$ and $F_{cb}$ vanish at the same time. Moreover, it is not difficult to see that we can even construct a local expression of one of the times in terms of the other by expanding the two functions $G_i$ around their maxima. As for the function $C_{ij}$ that would appear in the time component of the metric (to which we cannot apply the discussion of the previous section because both $F_{bc}$ and $F_{cb}$ now become zero), we get from its definition and that of $\Xi$ [see Eq. \eqref{Xi}] that
\begin{align}
\dfrac{1}{C_{ij}}=\left(1-\dfrac{\partial f_i}{\partial O_i}\dfrac{\partial O_i}{\partial \delta_i}\right)+\left(1-\dfrac{\partial f_j}{\partial O_j}\dfrac{\partial O_j}{\partial \delta_j}\right)\dfrac{F_{ji}}{F_{ij}}-F_{ji}.
\end{align}
Of the three terms on the right hand side, the first and the third ones have a well-defined limit on the surface where the maxima of $G_b$ and $G_c$ coincide. The second one, however, requires a more detailed study, since a direct evaluation leads to an indetermination. Given the form of this indetermination, the only way in which $1/C_{ij}$ might display a good behavior is that the limit $F_{ji}/F_{ij}$ be finite. In order to compute the limit of $1/C_{ij}$, we first expand $F_{bc}$ and $F_{cb}$ around their maxima. In the case of $F_{bc}$, the relevant expressions for this expansion are
\begin{align}
p_c&=\dfrac{1}{2}\gamma L_o \delta_c m\dfrac{1+(x_c^0)^2}{x_c^0}+\gamma L_o \delta_c m \dfrac{1-(x_c^0)^2}{x_c^0}(t_c-t_c^{\rm ext})+\mathcal{O}[(t_c-t_c^{\rm ext})^2],\\
\delta_cc&=2\arctan x_c^0-\dfrac{4x_c^0}{1+(x_c^0)^2}(t_c-t_c^{\rm ext})+\mathcal{O}[(t_c-t_c^{\rm ext})^2],\\
\cos\delta_cc&=\dfrac{1-(x_c^0)^2}{1+(x_c^0)^2}+\dfrac{8(x_c^0)^2}{[1+(x_c^0)^2]^2}(t_c-t_c^{\rm ext})+\mathcal{O}[(t_c-t_c^{\rm ext})^2],\\
\sin\delta_cc&=\dfrac{2x_c^0}{1+(x_c^0)^2}-\dfrac{4[1-(x_c^0)^2]}{[1+(x_c^0)^2]^2}(t_c-t_c^{\rm ext})+\mathcal{O}[(t_c-t_c^{\rm ext})^2],\end{align}
and so
\begin{align}
\delta_cc \cos\delta_cc-\sin\delta_cc&=\dfrac{2x_c^0}{1+(x_c^0)^2}\left[\arctan x_c^0\dfrac{1-(x_c^0)^2}{x_c^0}-1\right]+16\arctan x_c^0\dfrac{(x_c^0)^2}{[1+(x_c^0)^2]^2}(t_c-t_c^{\rm ext})+\mathcal{O}[(t_c-t_c^{\rm ext})^2],
\end{align}
where $\mathcal{O}(\cdot\,)$ denotes terms of an order equal to or higher than that of its argument. Then, recalling that $x_c^0$ verifies Eq. \eqref{x0}, 
\begin{align}
F_{bc}(t_c)=\left[\dfrac{8}{3}\arctan x_c^0\dfrac{x_c^0}{1+(x_c^0)^2} -2\dfrac{1-(x_c^0)^2}{1+(x_c^0)^2}\right](t_c-t_c^{\rm ext})+\mathcal{O}[(t_c-t_c^{\rm ext})^2],
\end{align}
where the coefficient of the leading order term has an approximate numerical value of 2.4651. In order to determine the limit of the quotient $F_{ji}/F_{ij}$, we still need to carry out a similar analysis for $F_{cb}$ and compare the orders of the leading terms. However, this cannot be achieved analytically because we lack a closed expression for the zeroes of $F_{cb}$ (i.e. the extrema of $G_b$). A numerical computation performed in Mathematica confirms that the dominant order in an expansion of $F_{cb}$ around its maximum is indeed linear in $t_b-t_{b,\rm ext}^{(1)}$, and reveals that the coefficient of this linear contribution depends on the value of the black hole mass $m$, unlike in the case of $F_{bc}$. This coefficient does not vanish for the studied range of masses since it is related to the value of the second derivative of $G_b$ at its maximum (indeed, they only differ in a sign) and we already know by numerical means that this extremum is a single zero of the first derivative. As a result, the factor $1/C_{ij}^2$ can be defined properly (actually, both for $i=b$ and $c$) at the point where the coincident maxima of $G_b$ and $G_c$ are reached. 

In the light of the results of this section, we conclude that a change of the origin of the angular time solves the problem that appeared to prevent a successful implementation of our formalism. Indeed, by means of a mass-dependent redefinition of the origin of the angular time, we have managed to match the images of the primitives $G_b$ and $G_c$ corresponding to the interior region, for any sufficiently large value of the mass. Taking into account this fact, and the good behavior of $F_{ij}$, $G_i$, and $C_{ij}$, we find no obstruction to attain a well-defined effective metric in the interior region between the surfaces interpreted as black and white horizons. The required global inversion of times can be constructed, e.g., in terms of three patches, expressing the angular time in terms of the radial one in each of these parts. In the first place, we can write a well-defined effective spacetime metric from the black horizon up to the transition surface. In the second place, another patch can cover a neighborhood of the coincident maxima, overlapping with the previous patch. Finally, a third patch would be necessary to describe the remaining piece of the interior region, i.e. from the vicinity of the coincident maxima to the white horizon. 

\section{Conclusions and discussion} \label{sec:conclusion}

Some few years ago, a new proposal for the effective description of black holes within the framework of LQC was put forward by the authors of Refs. \cite{AOS,AOS2,AO}. Their approach was based on the use of constants of motion to play the role of the polymerization parameters that introduce quantum effects in the system. They noticed that the model supplied naturally two constants of motion, namely the two partial Hamiltonians that generate the dynamics in the radial and angular sectors (two sectors of the phase space that are dynamically decoupled). The form of the Hamiltonian constraint implies that they are not only constant along dynamical trajectories, but also equal to each other. Their coincident value on the constraint surface is related to the mass of the black hole under consideration. The approach proposed in Refs. \cite{AOS,AOS2,AO} employs two polymerization parameters that are functions of this mass, corresponding to the on-shell value of the aforementioned constants of motion. The validity of this approach was supported with an argument involving an extension of the phase space (see Refs. \cite{AOS,AOS2}). The model resulting in this way displays a number of attractive features that make it stand out from previous related works. For instance, as far as the interior region is concerned, the classical central singularity is replaced with a transition surface that joins a trapped and an anti-trapped region, effectively extending the interior of a classical Schwarzschild black hole to a larger region bounded to the past by a black horizon and to the future by a white horizon. In this region, the effective spacetime metric is smooth and its curvature invariants are finite. Furthermore, unlike in the case of previous analyses, these results are independent of fiducial structures and (local) quantum effects appear to be confined to regions of large spacetime curvature. 

Despite these interesting properties, it has been pointed out that the model suffers form certain problems. Of particular relevance to the present article is an issue about the choice of polymerization parameters presented in Ref. \cite{N}. The authors of that work argue that the way in which these parameters were defined in the original model is inconsistent with the claim that they are constants of motion, so that the dynamical equations of the model have an unclear relation with the proposed effective Hamiltonian. Instead of taking the parameters as constant numbers determined by the value of the black hole mass, they propose to define each parameter as a function of its respective partial Hamiltonian, a procedure that leads to new terms in the equations of motion which are sourced by the resulting non-vanishing Poisson brackets of the polymerization parameters. The new equations differ from those considered previously in the fact that they include two additional phase space dependent factors that complicate the dynamics. In this context, in order to try and reconcile to some extent the results of the original model \cite{AOS,AOS2,AO} with a proper Hamiltonian treatment of the polymerization parameters, we recently put forward an alternative proposal that generalizes the approach of Ref. \cite{N}. We argued in Ref. \cite{AG} that, if both partial Hamiltonians have identical on-shell values, we should not be able to tell apart their contributions on the constraint surface. Thus, the most general choice of parameters should be such that each of them captures the contribution of both Hamiltonians, allowing a breaking of the decoupling of the radial and angular sectors. 

In Ref. \cite{AG} we limited our discussion to the introduction of the basic elements of our two-time formalism and to a preliminary analysis of the relation between the two times in the limit of infinitely large black hole masses, for regions where the limit is applicable. Nonetheless, we ignored the subtle issue of whether the formalism could be implemented without inconsistencies and lead to an effective metric that is well defined in the totality of the interior region in the first place. The aim of the present article is precisely to fill this gap, which is of crucial importance if we want to further examine the physical consequences of the model, setting our two-time formalism on firm grounds. Without this viability analysis, any future investigation of the features of the model would be meaningless, in the sense that one might even fail to have an acceptable effective geometry.

After a brief review of the main ingredients of our proposal in Sec. \ref{sec:basics}, we have proceeded to discuss whether there exist impediments to our description of the interior geometry in Sec. \ref{sec:obstructions}. The main aim of this section is the analysis of the implicit relation between the two times that arise as a direct result of our choice of polymerization parameters: the radial time and the angular time. Indeed, this relation is a fundamental piece to construct a well-defined spacetime geometry, with an effective metric that must be expressible in terms of a single time in every part of the interior region. For this to be possible, we need that certain functions $F_{ij}$ be integrable all over that region and that, at least, one of the resulting primitive functions $G_i$ be invertible around every point in this region. This local invertibility is a necessary condition, though still not sufficient, to pass from a two-time formalism to a single time in every patch used to describe the spacetime geometry. For a satisfactory single-time reformulation of the spacetime geometry, the images of the functions $G_i$ must match so as to allow that the whole interior region is covered by suitably combining local inversions. The integrability and local invertibility analysis has been carried out in Sec. \ref{subsec:integrabilityinvertibility}. We have shown that, since $F_{ij}$ are elementary functions defined in the whole interior region, their primitive functions $G_i$ exist. These primitives $G_i$ are invertible around every point of their domains except for their respective extrema: a single maximum in the case of  the angular primitive $G_c$, and a maximum and a minimum in the case of the radial one. For sufficiently large black hole masses and the same choice of time origin for the radial and angular sectors, we have proven that the values of $G_b$ and $G_c$ at their extrema do not coincide, so that the possible lack of a local inversion never affects the two primitive functions simultaneously. In Sec. \ref{subsec:imageG} we have then analyzed the images of the two functions $G_i$ and shown that they differ, implying the existence of a subregion where the equivalence relation between radial and angular times simply cannot be satisfied. This subregion contains the surroundings of the maximum of $G_b$. In that part of the interior we cannot express one of our times in terms of the other, and hence there is no way of obtaining a well-defined effective metric there. In practice, it seems possible to extend our formalism up to the transition surface, but not much farther beyond owing to the appearance of this obstruction. For completeness in our analysis, in Sec. \ref{subsec:analysisC} we have studied the factors $1/C_{ij}^2$ that appear in the time component of the effective metric. Potentially, these factors could also lead to singularities and/or degeneracies. It turns out that we can immediately rule out the presence of such singularities. In addition, the off-shell freedom that is present in our formalism is sufficient to provide factors $1/C_{ij}^2$ that remain different from zero in the evolution, leading to a non-degenerate behavior that could not be achieved with the more restricted proposal in Ref. \cite{N}. Finally, in Sec. \ref{sec:redefinitionorigins}, we have explored the possibility of shifting the origin of the angular time of the system, discussing whether and how this change can help in solving the problems found in the formalism. A change in that time origin modifies the relation between the two times by introducing an additive constant in the equality between the functions $G_b$ and $G_c$. Then, a suitable fixation of this constant allows us to match the upper endpoints of the images of these functions. Therefore, the obstruction that we had found to apply our formalism to the whole interior region can be circumvented. As a result, we see no impediment to obtain a well-defined effective spacetime metric in the interior region with the alternative model that follows from our proposal to define the polymerization parameters. 

In conclusion, the proposal for the definition of the polymerization parameters put forward in Ref. \cite{AG} seems to lead to a viable effective description of the interior region of non-rotating, uncharged black holes. We have shown that the considered formalism allows us to cover the whole interior with (three) separate patches where a well-defined single-time effective metric can be constructed. The additional off-shell freedom that results from considering a more general choice of polymerization parameters makes it possible to avoid the singularities that inevitably appeared in other previous two-time models \cite{N}. Once we have proven that, in principle, the obstructions to our effective model of the interior geometry are solvable, we have the necessary groundwork to explore its physical properties. For this, the next logical step would be to investigate the causal structure of the effective geometry and extend our proposal to the exterior region. We plan to address these issues in future works. 

It is worth recalling that our main motivation to study these alternatives is to reconcile the results of the original AOS model with a more standard treatment of the parameters as constants of motion, with an eye on finding a self-consistent Hamiltonian formalism which one could proceed to quantize. In this regard, it seems to us that other open possibilities exist. For instance, the question stands of whether we can find another route to a consistent Hamiltonian formulation that leads exactly to the dynamical equations of the AOS model, for which the interior spacetime geometry is well defined and displays some nice properties. Certainly, one may follow the suggestion of the authors of Refs. \cite{AOS,AOS2,AO} and handle the polymerization parameters as constant numbers in the derivation of the Hamiltonian equations, evaluating them as constants of motion only after the calculation has been done. This parachuting provides the desired dynamics, and hence leads to the AOS solution, but would be debatable from a quantum perspective, since the result of the quantization would be different for a Hamiltonian in which the polymerization parameters are either $c$-numbers or Dirac constants. Thus, it may be enlightening to consider as another option the extended phase formalism proposed in Ref. \cite{AOS2}, exploring whether a suitable reduction may lead both to the desired dynamics and to parameters that are indeed manifest constants of motion in the system. This matter will constitute the subject of future research. 

\acknowledgments
	
The authors are grateful to B. Elizaga Navascu\'es for fruitful discussions, as well as for enlightening comments on earlier versions of this manuscript. This work has been supported by Project. No. MICINN PID2020-118159GB-C41. The project that gave rise to these results received the support of a fellowship from ``la Caixa'' Foundation (ID 100010434). The fellowship code is LCF/BQ/DR19/11740028.

\end{document}